# Invisible Manipulation: Deep Reinforcement Learning-Enhanced Stealthy Attacks on Battery Energy Management Systems

Qi Xiao, *Student Member, IEEE*, Lidong Song, *Member, IEEE*, Jongha Woo, *Student Member, IEEE*, Rongxing Hu, *Member, IEEE*, Bei Xu, *Member, IEEE*, Kai Ye, *Student Member, IEEE*, Ning Lu, *Fellow, IEEE*

*Abstract*—This paper introduces "invisible manipulation," an innovative cyber-attack mechanism achieved through strategically timed stealthy false data injection attacks (SFDIAs). By stealthily manipulating measurements of a critical asset prior to the target time period, the attacker can subtly guide the engineering system toward a predetermined operational state without detection. Using the battery energy management system (BEMS) as a case study, we employ deep reinforcement learning (DRL) to generate synthetic measurements, such as battery voltage and current, that align closely with actual measurements. These synthetic measurements, falling within the acceptable error margin of residual-based bad data detection algorithm provided by state estimation, can evade detection and mislead Extended Kalman-filter-based State of Charge estimation. Subsequently, considering the deceptive data as valid inputs, the BEMS will operate the BESS towards the attacker-desired operational states when the targeted time period come. The use of the DRL-based scheme allows us to covert an online optimization problem into an offline training process, thereby alleviating the computational burden for real-time implementation. Comprehensive testing on a high-fidelity microgrid real-time simulation testbed validates the effectiveness and adaptability of the proposed methods in achieving different attack objectives.

*Index Terms*—Cyber-physical attacks, deep reinforcement learning, delayed stealthy false data injection (SFDIA), invisible manipulation attacks, state-of-charge (SoC) estimation.

## I. Introduction

THE secure and reliable operation of the electric grid heavily relies on accurate data for decision-making and control, making false data injection attacks (FDIAs) a significant threat. Over the past decade, research has focused on FDIAs targeting transmission systems by falsifying inputs to state estimation (SE) algorithms and evading detection by bad data detection (BDD) algorithms (e.g., residual-based BDD) [1]. With the rapid integration of distributed energy resources (DERs) [2] and other information and communication technology devices (ICTs), previously passive distribution networks are evolving into smart grids, equipped with numerous remotely accessible automated devices [3]. Consequently, the risk posed by FDIAs to modern active distribution networks (ADNs) has been growing exponentially. This has led to an increased emphasis on studying FDIAs in the context of DERs [4].

Battery energy storage system (BESS) plays a critical role in many ADNs by providing grid-following and grid-forming functions. These functions include PV output smoothing, load shifting, and voltage and frequency regulation [5], [6].

This research is supported by the U.S. Department of Energy's Office of Energy Efficiency and Renewable Energy (EERE) under the Solar Energy Technologies Office Award Number DE-EE0008770. The authors are with the Department of Electrical and Computer Engineering, North Carolina State University, Raleigh, NC, 27606 USA. E-mails: (qxiao3@ncsu.edu and nlu2@ncsu.edu).

However, the versatility and significance of BESS also make it prime target for adversaries seeking to launch FDIAs and exploit them for unlawful purposes.

As presented in Table I, FDIA objectives targeting BESS can be categorized into two main types: manipulation of control parameters within inverters (e.g., real and reactive power, voltage, and frequency measurements) and tampering with the measurement and status data of batteries (e.g., state of charge (SoC)). In this paper, we focus on the latter due to the practical challenges of directly measuring SoC with sensors. SoC estimation relies on techniques such as Coulomb counting or Kalman filter-based methods, which use available battery measurements from the local battery management system (BMS) [14]. Therefore, the accuracy of SoC estimation depends heavily on these measurements, making it vulnerable to cyber-attacks. This vulnerability is exacerbated by the growing dependence on remote communication between BESS and the control center in today's ADNs. The integration of communication-based technologies like IoT and cloud computing to streamline SoC estimation in BMS [15] further exposes BESS to manipulation through sensor faults or falsified measurements.

Additionally, both BESS energy scheduling and battery operation lifespan depend on accurate SoC estimation. Corrupted SoC measurements can misguide the Battery Energy Management Systems (BEMS) in power systems, leading to erroneous decisions, insufficient support for ongoing operations, or potential future energy shortages. Moreover, incorrect SoC values can result in overcharging, over-discharging, reduced battery lifecycles, and serious risks such as fire or explosion [14].

Thus, we introduce a novel FDIA scheme designed to manipulate SoC estimation invisibly, misleading the BEMS to create incorrect dispatch plans, thereby compromising the BESS's ability to provide effective grid support within the target period.

Based on attack duration, FDIAs are categorized into *instantaneous* and *prolonged* attacks. Prolonged attacks are further classified into *persistent* attacks, where a constant bias is permanently added to authentic data, and *repetitive* attacks, where different bias values are periodically introduced [10]. Existing research on FDIAs against BESS has mainly focused on their impact on power grid stability through persistent and repetitive attacks, as shown in Table I with rule-based FDIAs. However, the technical implementation details of these attacks have not been extensively discussed.

Rule-based FDIAs involve injecting false data randomly within the operational range, causing persistent or repetitive biases in measurements and commands. These attacks are relatively straightforward but can be detected through personnel observations and BDD mechanisms, which are commonly used to identify and mitigate cyber-attacks or inaccuracies. In contrast, stealthy FDIAs (SFDIAs), which can evade certain BDD mechanisms, pose an even greater threat.

TABLE I
LITERATURE REVIEW OF EXISTING FDIAS TARGETING BESS

| Type | Attack Objectives | BDD (Y/N) | Description | Advantages | Disadvantages |
|---|---|---|---|---|---|
| Rule-based attacks | Power control [7] | No | Inject bias within the operation range to the active power setpoints of BESS to cause power imbalance in an islanded microgrid. | 1. Conscious manipulation. 2. Easy to implement. | 1. Non-stealthy. 2. Easy to be detected by personnel observations or BDD. |
| | Mode control [8] | | Falsify the mode command to disrupt the mode conversion from PQ to Vf to fail the microgrid. | | |
| | ON/OFF control [9] | | Falsify the ON/OFF command to deteriorate power quality or destabilize the power system. | | |
| | SoC estimation [10] | | Different voltage bias is selected within the operation range to disrupt the SoC estimation. | | |
| Optimization-based attacks | SoC estimation [11] | Yes | Maximize the SoC estimation error to cause overcharging or over-discharging of batteries with residual-based BDD considered. | 1. Highly stealthy. 2. Maximize the absolute SoC estimation error. | 1. System information is required. 2. Many simplifications and assumptions are made. 3. High computation cost and compromised accuracy, not applicable for real-time attacks. |
| Machine learning-based attacks | BESS operation status [12],[13] | No | Utilize an artificial neural network to replicate the normal behavior of BESS for enhanced stealth and the attacker takes control of the authentic BESS by employing MitM techniques. | 1. Mediumly stealthy. 2. Conscious manipulation. | 1. Falsify a large amount of data. 2. Easy to be detected by residual-based BDD. |
| | **SoC estimation (Proposed)** | Yes | DRL-based SoC estimation attack scheme to introduce a maximum or target SoC error at the desired time by gradually injecting battery voltage and current measurement bias. | 1. Highly stealthy. 2. High modeling accuracy in real-time applications. 3. Conscious manipulation of SoC estimation error. | 1. System information is required. 2. Offline training is required. |

State estimation (SE), particularly residual-based BDD, is commonly used for identifying cyberattacks and measurement errors [16]. In ADNs, distribution network SE can incorporate measurements and states from inverter-based DERs. Previous studies introduced SFDIA architectures for BESS, utilizing man-in-the-middle (MitM) attacks to manipulate commands and measurements between the BESS controller and BMS [14], [15]. These architectures ensure stealth at the local BESS controller level by utilizing artificial neural networks to replicate normal BESS behavior, allowing the attacker to control the authentic BESS. However, these studies did not consider evading detections, such as residual-based BDD, making them susceptible to detection at the system network level.

Recent research highlights vulnerabilities in BDD algorithms, allowing certain FDIAs to evade detection, including SFDIAs. For instance, [11] proposed an SFDIA scheme to maximize the absolute SoC estimation error in BESS by integrating the residual-based BDD mechanism with linear optimization methods. However, this approach encountered challenges due to the need for numerous linearization and assumptions to simplify problem formulation and reduce computation costs. These assumptions, such as considering the slope of the OCV-SoC curve as a constant in the Extended Kalman Filter (EKF), adopting linearized and DC power flow instead of an AC power flow model, and assuming system data for future time slot are available, made real-time implementation difficult and increased susceptibility to detection in practical scenarios [17], [18].

In contrast, deep reinforcement learning (DRL) shows promise for launching real-time attacks, offering an accurate system model through offline training. DRL has been applied in power systems for decision-making and control [19], [20], with recent developments focusing on enhancing cyber security [21], [22]. While DRL is often utilized for attack detection, it can also be a potent tool for launching cyber-attacks. This study leverages DRL to orchestrate stealthy attacks on SoC estimation within BESS, aiming to bridge the gap in understanding the attack design methodology integrated with BEMS. Different from simply injecting the constant bias in both persistent and repetitive attacks, we define a delayed attack in prolonged attacks that features injecting a different small bias at each time step to accumulate the error and finally reach the desired error at the specific time points.

The primary contribution of this paper is the development of a DRL-based approach for launching SFDIAs against SoC estimation. Unlike optimization-based methods, which require online optimization, substantial computing resources, and extended runtimes, the DRL-based approach allows for offline training of an attacker agent within a simulated environment. This method uses accurate models, such as the AC power flow model and precise SoC estimation, without simplifying the problem formulation, enhancing response accuracy.

A secondary contribution is the introduction of a delayed attack scheme with a specific SoC error target. By strategically falsifying SoC well in advance of the targeted attack time, we incrementally introduce SoC errors between false and actual values. This gradual approach mitigates detection risk while guiding the BEMS to operate the BESS at attacker-specified SoC levels when the target period arrives. An attack scenario named "one-shot kill" exemplifies the delayed attack scheme, aiming to undermine the reliability of BESS-supplied systems.

The effectiveness of the proposed DRL attack techniques and delayed attack schemes is validated through rigorous testing on a real-time HIL simulation platform, with simulation results affirming the efficacy of the proposed methods.

The rest of this paper is structured as follows: Section II

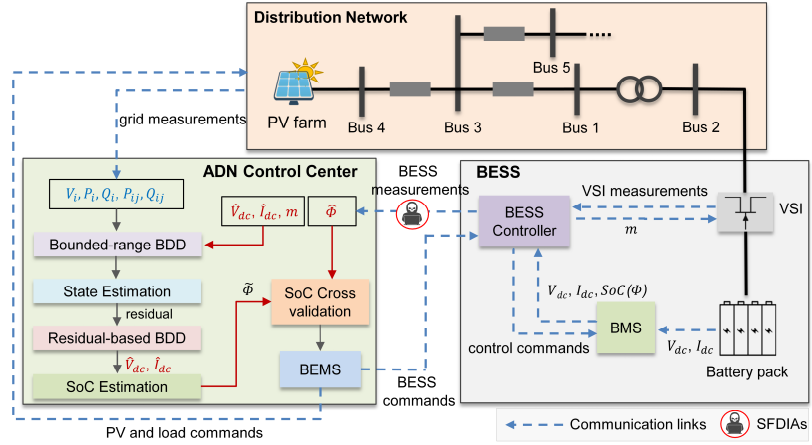

**Fig. 1.** A single-line diagram of an ADN with BESS.

outlines the ADN configuration with BESS and the BDD mechanism. Section III formulates the DRL-based delayed SFDIA issue and discusses the design methodology. Section IV presents case study results from implementing the proposed attack scheme on a high-fidelity microgrid real-time simulation testbed. Section V concludes the paper.

## II. MODELING CONSIDERATIONS

This section provides an overview of the ADN system configuration, the BDD mechanism at ADN control center, and the proposed delayed SFDIA scheme.

### A. Configuration of the ADN System

As depicted in the grey box in Fig. 1, the customer-owned BESS consists of a bidirectional three-phase voltage source inverter (VSI), a battery pack, a battery management system (BMS), and a local BESS controller. The BMS uses local measurements from the battery pack, including the BESS DC terminal voltage ($V_{dc}$) and current ($I_{dc}$), to estimate the BESS SoC ($\Phi$) and other relevant operation variables. The BESS controller receives the measurement data from VSI and BMS, determining the inverter modulation index ($m$) of the VSI. Then, it transmits $V_{dc}, I_{dc}, m$, and $\Phi$ to the ADN control center. These data are transmitted via the internet or non-proprietary wireless networks, where the risk of injecting fake data exists. For clarity and convenience, we define the BESS measurement set sent to the ADN control center as

$$\boldsymbol{z}_{BESS} = [V_{dc}, I_{dc}, m, \Phi] \quad (1)$$

Using the average model introduced in [23] to represent the VSI operation of the BESS unit, we have

$$V_r = mV_{dc}/\sqrt{2} \quad (2)$$
$$P_{ri} + P_{dc} + P_{loss} = 0 \quad (3)$$
$$P_{dc} = V_{dc}I_{dc} \quad (4)$$
$$P_{loss} = I_{ri}^2 R_{ac} + V_{dc}^2/R_{dc} \quad (5)$$

where $P_{dc}$ represents the DC power, $P_{ri}$ and $I_{ri}$ denote AC side power and current, $P_{loss}$ signifies inverter loss, and $R_{ac}, R_{dc}$ are AC and DC bus series resistances, respectively.

As illustrated in the green box in Fig. 1, in addition to retrieving BESS operation status from the BESS controller, the ADN control center collects measurements from the distribution network through proprietary Supervisory Control and Data Acquisition (SCADA) systems. These measurements are crucial for SE, a process that processes a set of noisy and redundant measurement data to provide an accurate real-time database for control and monitoring purposes [16]. The SCADA measurement set include vital operation parameters, such as voltage ($V_i$), phase angle ($\theta_i$), active and reactive power injections at the $i^{th}$ bus ($P_i, Q_i$), as well as active and reactive power flows between buses $i$ and $j$ ($P_{ij}, Q_{ij}$). This measurement set from SCADA is defined as

$$\boldsymbol{z}_{SCADA} = [V_i, \theta_i, P_i, Q_i, P_{iJ}, Q_{ij}, P_i] \quad (6)$$

### B. Bad Data Detection

To ensure the reliability of measurements and mitigate the impact of potential cyber-attacks, this paper assumes the implementation of three distinct BDD algorithms at the ADN control center: bounded-range BDD, residual-based BDD, and SoC cross-validation. Fig. 2 provides a flowchart illustrating the operation of these BDD mechanisms within ADN control center.

#### 1) Bounded-range BDD

Bounded-range BDD involves threshold-based checks applied to all incoming measurements upon receipt by the ADN control center. These checks compare the received measurements against predefined upper and lower bounds, as defined by equations (7) ~ (8) respectively. If any measurement falls outside these bounds, an alarm is triggered, signaling potential data anomalies or attacks.

$$\boldsymbol{z}_{SCADA}^{min} \leq \boldsymbol{z}_{SCADA} \leq \boldsymbol{z}_{SCADA}^{max} \quad (7)$$
$$\boldsymbol{z}_{BESS}^{min} \leq \boldsymbol{z}_{BESS} \leq \boldsymbol{z}_{BESS}^{max} \quad (8)$$

#### 2) Residual-based BDD

Residual-based BDD is employed subsequent to bounded-range BDD. This mechanism evaluates the validity of measurements by comparing the SE residual ($\bar{r}$) with a predefined threshold value ($\tau_{SE}$). If both BESS and SCADA measurements pass the bounded-range BDD, the SE estimates system states ($\boldsymbol{x}$) that best match the measurements ($\boldsymbol{z}$) via:

$$\boldsymbol{z} = \boldsymbol{h}(\boldsymbol{x}) + \boldsymbol{e} \quad (9)$$

Here, $\boldsymbol{h}(\cdot)$ denotes a nonlinear vector function derived from network topology and $\boldsymbol{e}$ is the measurement error vector.

Using the Weighted Least Squares (WLS) method, we minimize weighted measurement residuals and iteratively solve the optimization problem detailed in [24]. Thus, the residual-

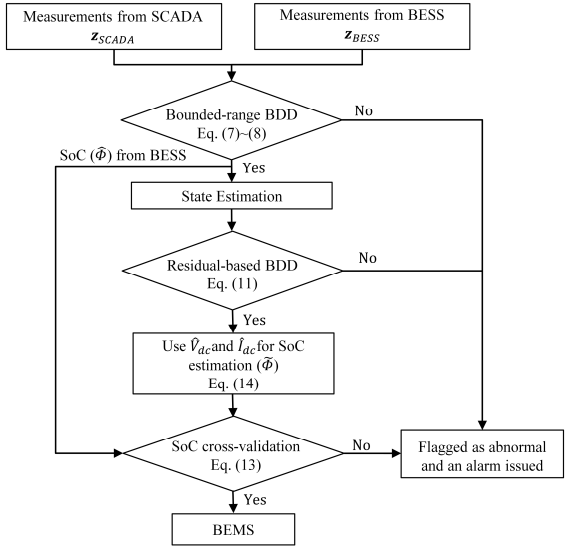

**Fig. 2.** Flow chart of bad data detection at the ADN control center.

based BDD mechanism validates measurements based on residual, $\bar{r}$, calculated as

$$\bar{r} = \|z - h(\hat{x})\|_2^2 \qquad (10)$$

where $\hat{x}$ represents the estimated value of $x$.

If $\bar{r} \leq \tau_{SE}$, $z$ is considered as normal; otherwise, $z$ is flagged as containing bad data. Thus, a SFDIA that can pass the residual-based BDD, needs to generate a set of measurement attack biases $\varepsilon$ that can satisfy

$$\|z + \varepsilon - h(\hat{x} + c)\|_2^2 \leq \tau_{SE} \qquad (11)$$

where $c$ represents the malicious error introduced to the original estimation $\hat{x}$.

Many existing methods assume that a large number of measurements in $z$ can be modified. For example, in [11], it is required to alter not only the battery voltage and current but also other SCADA measurements to maintain the residual consistency between attacked and non-attacked cases. This assumption is restrictive, as it necessitates the attacker having access to a lot of communication links. Consequently, in this paper, we investigated SFDIA under a more realistic scenario where the attacker can only manipulate the battery measurements by adding an attack vector $\varepsilon_{BESS1}$. Under this assumption, (11) becomes

$$\|z_{SCADA} + z_{BESS1} + \varepsilon_{BESS1} - h(\hat{x} + c)\|_2^2 \leq \tau_{SE} \qquad (12)$$

where $z_{BESS1} = [V_{dc}, I_{dc}, m]$ and $\varepsilon_{BESS1}$ is the attack vector containing the biases of battery voltage and current. Please note SoC cannot be directly included for SE. Thus, SoC is excluded in $z_{BESS1}$ compared to $z_{BESS}$.

*3) SoC Estimation and Cross-validation*

Given the critical role of BESS SoC in system operation, accurate estimation of SoC is essential. The battery voltage and current passing residual-based BDD are used to calculate an estimated SoC ($\widetilde{\Phi}$) in the ADN control center, cross-validated against the received SoC ($\Phi$), as shown in Fig. 1. Assuming the ADN control center and the local BMS use the same battery model for SoC estimation. If the discrepancy between $\widetilde{\Phi}$ and $\Phi$ is within the threshold $\tau_{SoC}$, as shown in (13), the received SoC $\Phi$ is deemed valid and incorporated into the BEMS for battery operation scheduling. Otherwise, an alarm will be issued.

$$|\widetilde{\Phi} - \Phi| \leq \tau_{SoC} \qquad (13)$$

The Extended Kalman filter (EKF) is a widely adopted method for SoC estimation due to its capability to manage measurement errors, noise, and initial value uncertainties. This paper employs the EKF-based SoC estimation process introduced in [14], represented by:

$$\Phi = EKF(V_{dc}, I_{dc}) \qquad (14)$$

*C. Proposed Delayed SFDIA Scheme*

To effectively launch an SFDIA targeting SoC estimation while bypassing existing BDD mechanisms, it is necessary to at least falsify three key measurements: SoC, battery voltage, and current. This paper introduces a delayed SFDIA scheme named "one-shot kill", which gradually introduces SoC estimation errors, ultimately leading to the shutdown of the BESS-supplied system. An illustrative example of the one-shot kill scheme involves inducing a microgrid blackout by misleading the BEMS into believing the battery SoC is sufficient to support the microgrid throughout the night. In reality, the SoC falls below critical levels after midnight.

Figure 3 illustrates one of the scenarios, where the BEMS targets maintaining SoC around 90% by hour 18 and 45% by hour 24 to ensure microgrid stability. However, if the tampered SoC readings suggest compliance while actual SoC levels drop to 70% by hour 18 and 20% by hour 24, the microgrid may face shutdown due to insufficient energy storage.

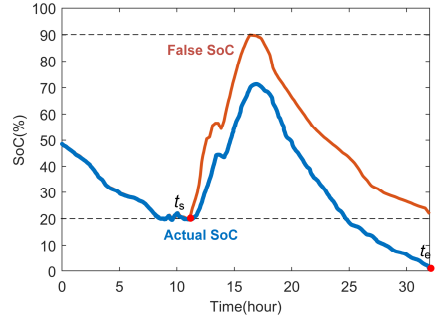

**Fig. 3.** Illustration of one-shot kill attack.

The advantage of delayed attacks lies in its ability to initiate false data injections at an earlier time $t_s$, significantly before the intended target time $t_e$. This strategy gives the attacker ample opportunity to discreetly modify the data stream, thereby avoiding detection by the BDD process. As a result, when the attack is executed, the alterations are less likely to be flagged.

To achieve the one-shot kill, we propose two attack methods: *unconstrained* and *constrained* SoC error attack. The unconstrained attack aims to maximize SoC error by $t_e$, where the exact error is unknown. Conversely, the constrained SoC error attack aims to inject a specific, desired SoC error by the end of attack. The attack process of both attacks is at each timestep from $t_s$ to $t_e$, a bias $\varepsilon_{BESS1}$ containing battery voltage and current biases is injected into $z_{BESS1}$. Based on these injected biases, a false SoC is estimated to replace the actual SoC. At each timestep, the altered data must satisfy the three BDD constraints, as outlined in (8), (12), and (13).

The inherent complexity and nonlinearity of delayed stealthy attacks, compounded by the EKF-based SoC estimator, pose challenges that require computationally intensive optimization. This optimization may necessitate accurate future system information for optimal performance. Therefore, in the following section, we propose a DRL-based method to address these complexities while maintaining system accuracy.

## III. METHODOLOGY

### A. Proposed DRL Framework for Delayed SFDIAs

Reinforcement learning involves an agent interacting with an environment, maximizing cumulative rewards to learn the best action based on different states. This process can be modeled as a Markov Decision Process (MDP), where the future depends solely on the present state [25]. The MDP can be represented as a tuple $\{S, A, P, R, \gamma\}$, where $S$ is the environment state space, $A$ is the action space, $P$ denotes the transition probability, $R$ is the reward function and $\gamma \in [0, 1]$ denotes the discount rate for the long-term return.

In this paper, we formulate delayed SFDIAs against SoC estimation as a MDP and use an agent to emulate the attacker. Figure 4 illustrates the proposed actor-critic-based DRL framework for delayed SFDIAs against SoC estimation. The agent consists of an actor and critics, and the critic evaluates the performance of actions generated by the actors during training. Here, the global observation $o_t$, state $s_t$, action set $a_t$ and attack vector $\varepsilon_{BESS}^t$ are defined as follows:

$$o_t = [z_{SCADA}^t, z_{BESS1}^t] \tag{15}$$

$$s_t = \left[z_{SCADA}^t, z_{BESS1}^t, \varepsilon_{BESS}^{t-1}, \frac{t}{T}\right] \tag{16}$$

$$a_t = [\Delta V_{dc}^t, \Delta I_{dc}^t] \tag{17}$$

$$\varepsilon_{BESS}^t = [\varepsilon_{BESS1}^t, \varepsilon_{BESS2}^t] \tag{18}$$

$$\varepsilon_{BESS1}^t = [\sum_{t_s}^t \Delta V_{dc}^t, \sum_{t_s}^t \Delta I_{dc}^t] \tag{19}$$

$$\varepsilon_{BESS2}^t = [\Delta \Phi^t] \tag{20}$$

where $o_t$ includes the measurements from SCADA and BESS at time $t$; $s_t$ includes these measurements, the BESS attack vector taken at time $t-1$, and the ratio of the current timestep $t$ to the total attack duration $T$; $a_t$ includes the battery voltage and current bias at time $t$; $\varepsilon_{BESS}^t$ comprises the accumulated battery voltage and current bias vector $\varepsilon_{BESS1}^t$ by time $t$, and the current SoC error $\Delta \Phi^t$ between the false SoC and actual SoC in $\varepsilon_{BESS2}^t$. Note that $a_{t-1} = 0$ when $t = 1$.

The attacker's actions involve injecting the attack vector $\varepsilon_{BESS}^t$ into the measurement from the BESS. These biases manipulate the battery DC voltage, current, and SoC data streams sent to the ADN controller (as shown in Fig. 1). The manipulated battery voltage, current and SoC at time $t$ are denoted as $\hat{V}_{dc}^t$, $\hat{I}_{dc}^t$ and $\hat{\Phi}^t$, respectively.

As illustrated in Fig. 4, at each timestep $t$, the attacker agent observes the system state $s_t$, and the actor network generates an action $a_t$ based on this state. We assume the attacker also knows the parameters of the EKF-based SoC estimator. Therefore, instead of generating the SoC bias directly from the actor network, the fake SoC $\hat{\Phi}^t$ at time $t$ is first estimated using the false voltage $\hat{V}_{dc}^t$ and current $\hat{I}_{dc}^t$, as shown in equations (21) to (23). Here, $V_{dc}^t$ and $I_{dc}^t$ are the actual voltage and current data at time $t$. The SoC bias $\Delta \Phi^t$ in $\varepsilon_{BESS2}^t$, defined in (24), is calculated by comparing the false SoC $\hat{\Phi}^t$ with the actual SoC $\Phi^t$. The actual SoC is represented in (25).

$$\hat{\Phi}^t = EKF(\hat{V}_{dc}^t, \hat{I}_{dc}^t) \tag{21}$$

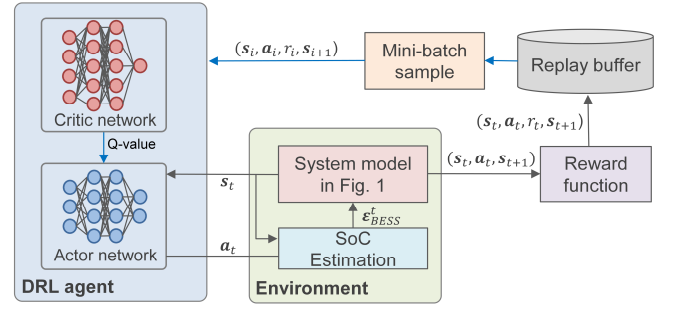

**Fig. 4.** Proposed actor-critic-based DRL framework for delayed SFDIAs against SoC estimation.

$$\hat{V}_{dc}^t = V_{dc}^t + \sum_{t_s}^t \Delta V_{dc}^t \tag{22}$$

$$\hat{I}_{dc}^t = I_{dc}^t + \sum_{t_s}^t \Delta I_{dc}^t \tag{23}$$

$$\Delta \Phi^t = \hat{\Phi}^t - \Phi^t \tag{24}$$

$$\Phi^t = EKF(V_{dc}^t, I_{dc}^t) \tag{25}$$

Inputting the attack vector into the system model, the system transitions to a new state $s_{t+1}$ at the next timestep, generating a reward $r_t$. The tuple $(s_t, a_t, r_t, s_{t+1})$ for each timestep is stored in the replay buffer. During the training stage, a batch of data is randomly sampled to update the parameters of critic and actor neural networks once enough data has been collected.

### B. Reward Function Design

The key challenge of delayed SFDIAs is to generate a sequence of attack vectors $a_t$ that consistently evades BDD at each time step, ensuring that the desired SoC error is achieved by the end of attack period. This paper presents two reward mechanisms: unconstrained and constrained.

The objective of unconstrained attack is to maximize the SoC error at the end of attack $t_e$. To achieve this, if there is no BDD violation, the agent receives a reward $r_{u1,t}$ at each time step starting from $t_s$, based on the current injected SoC error. If, during the attack, any attack vector violates the desired system operation range (as specified in equations (26) to (28)), resulting in a failure to pass any of the bounded-range, residual-based and SoC cross-validation BDDs, a penalty $p_u$ with a large negative value is applied to the reward. At the end of attack $t_e$, an additional bonus $r_{u2}$ is given.

$$[V_{dc}^{min}, I_{dc}^{min}, \Phi^{min}] \leq [\hat{V}_{dc}^t, \hat{I}_{dc}^t, \hat{\Phi}^t] \leq [V_{dc}^{max}, I_{dc}^{max}, \Phi^{max}] \tag{26}$$

$$\|z_{SCADA}^t + z_{BESS1}^t + \varepsilon_{BESS1}^t - h(\hat{x}^t + c^t)\|_2^2 \leq \tau_{SE} \tag{27}$$

$$|\tilde{\Phi}^t - \hat{\Phi}^t| \leq \tau_{SoC} \tag{28}$$

The reward function ($r_t$) for the unconstrained SoC error attack are:

$$r_t = r_{u1,t} + r_{u2} + p_u \tag{29}$$

$$r_{u1,t} = \frac{\Delta \Phi^t}{k_{u2}} \times k_{u1}, \quad t \in [t_s, t_e] \tag{30}$$

$$r_{u2} = \begin{cases} 0, & t \in [t_s, t_e) \\ r_{u1,t} \times k_{u3}, & t = t_e \end{cases} \tag{31}$$

$$p_u = k_p \times F_{bdd}, \quad t \in [t_s, t_e] \tag{32}$$

In equation (30), $k_{u1}$ denotes the sign of the desired SoC error: +1 indicates the false SoC is intended to be higher than the actual SoC, and -1 indicates lower. The coefficient $k_{u2}$ is used as a normalization factor for $r_{u1,t}$. Equation (31) offers a bonus based on final SoC error to incentivize larger errors. In

(32), $k_p$ specifies the magnitude of the penalty, with $F_{bdd}$ flagging BDD violation (1 for violations, 0 otherwise).

The objective of the constrained SoC error attack is to achieve and maintain a targeted SoC error ($\Delta\Phi^*$) by the end of attack. This means, once the $\Delta\Phi^*$ is reached, there is no need to inject higher SoC errors, risking being detected. Then, the reward function is adjusted as follows:

$$r_t = r_{tar1,t} + r_{tar2} + p_u \quad (33)$$

$$r_{tar1,t} = min\left\{2 - \frac{\Delta\Phi^t}{\Delta\Phi^*}, \frac{\Delta\Phi^t}{\Delta\Phi^*}\right\} \times k_{t1}, t \in [t_s, t_e] \quad (34)$$

$$r_{tar2} = \begin{cases} 0, & t \in [t_s, t_d) \text{ or } (t_d, t_e) \\ r_{tar1,t} \times k_{t2} + r_{tar3}, & t = t_d \text{ or } t_e \end{cases} \quad (35)$$

$$r_{tar3} = \begin{cases} 0, & |\Delta\Phi^t - \Delta\Phi^*| > 1\% \\ k_{t3}, & |\Delta\Phi^t - \Delta\Phi^*| \leq 1\% \end{cases} \quad (36)$$

As shown in equation (34), the reward received at each time step $r_{tar1,t}$ is associated with how close the current SoC error $\Delta\Phi^t$ is towards the target $\Delta\Phi^*$. We use *min{}* to limit its maximum value to 1. Thus, the maximum value is only achieved when the injected SoC error $\Delta\Phi^t$ equals the target $\Delta\Phi^*$. $k_{t1}$ is a coefficient to adjust the value of $r_{tar1,t}$.

The sign of the target $\Delta\Phi^*$ determines whether the false SoC $\hat{\Phi}^t$ is higher or lower than the actual SoC $\Phi^t$. Specifically, if $\Delta\Phi^* > 0$, $\hat{\Phi}^t$ is higher than $\Phi^t$ by $\Delta\Phi^*$ at $t_e$; Conversely, if $\Delta\Phi^* < 0$, $\hat{\Phi}^t$ is lower than $\Phi^t$ by $\Delta\Phi^*$ at $t_e$.

To motivate the agent to achieve the targeted SoC error, an additional bonus $r_{tar2}$ in (35) is given at two specific points: the designated time $t_d$ and the end time $t_e$. The highest reward is given if the absolute difference between the injected SoC error and the targeted value is within 1% of SoC at these points. $k_{t2}$ and $k_{t3}$ are the coefficients to adjust the bonus value. The dual bonus system ensures that the agent keeps the SoC error until the end of the attack, if the target is reached at $t_d$.

Balancing penalties and rewards are crucial for optimal agent performance. Excessive penalties may cause inaction to avoid BDD violations, while overly high rewards could lead to frequent BDD violations for higher returns. Additionally, balancing the attack-end bonus and cumulative step-by-step rewards is necessary to achieve the desired SoC error at specific attack times.

*C. Soft Actor-Critic (SAC) Framework*

Based on this reward mechanism, the goal of the delayed SFDIA issue is to maximize the sum of the expected discounted rewards over the attack horizon of $T$:

$$\max_\pi J = \mathop{E}_{(s_t, a_t) \sim \pi}\left[\sum_{t=0}^{T}(\gamma)^t \cdot r_t(s_t, a_t)\right] \quad (37)$$

where $E(\cdot)$ represents the mathematical expectation, $\pi$ is the actor policy that generates action according to state $s_t$, $r_t(s_t, a_t)$ is the reward (equation (26) or (33)) based on current state $s_t$ and action $a_t$. In this paper, we employ soft actor-critics (SAC) algorithm in [26] to find the optimal policy.

SAC is a model-free, off-policy actor-critic algorithm [27]. Unlike traditional actor-critic methods, SAC maximizes cumulative rewards and policy entropy, enhancing stochastic exploration and improving convergence speed and optimization performance. The objective function in equation (37) is transformed to equation (38) using SAC.

$$\pi^* = \arg\max_\pi \mathop{E}_{(s_t, a_t) \sim \pi}\left[\sum_{t=0}^{T}(\gamma)^t \left(r_t(s_t, a_t) + \alpha H(\pi(\cdot|s_t))\right)\right] (38)$$

where $\pi^*$ represents the optimal policy, $H(\pi(\cdot|s_t)) = -\log(\pi(\cdot|s_t))$ is the policy entropy, and $\alpha$ is the temperature parameter balancing entropy and reward.

In SAC, policy evaluation and improvement are achieved via training deep neural networks using stochastic gradient descent. SAC employs two networks: the $Q$ network $Q_\theta(s_t, a_t)$ approximates the state-action value function, and the policy network $\pi_\phi(a_t|s_t)$ approximates the policy function. The $Q$ network parameters $\theta$ are trained by minimizing the soft Bellman residual:

$$J_Q(\theta) = \mathop{E}_{(s_t, a_t) \sim D}\left[\frac{1}{2}\left(Q_\theta(s_t, a_t) - \left(r_t(s_t, a_t) + \gamma \mathop{E}_{s_{t+1} \sim p}[V_{\bar{\theta}}(s_{t+1})]\right)\right)^2\right] \quad (39)$$

where $V_{\bar{\theta}}(s_{t+1})$ is the estimated soft state value using a target network updated via moving average.

For continuous action spaces, the policy is modeled as a Gaussian distribution. The policy network outputs the mean and standard deviation of the action distribution. Actor network parameters $\phi$ are learned by minimizing the expected Kullback-Leibler divergence: [26]:

$$J_\pi(\phi) = \mathop{E}_{s_t \sim D, a_t \sim \pi_\phi}\left[\alpha\log\left(\pi_\phi(a_t|s_t)\right) - Q_\theta(s_t, a_t)\right] \quad (40)$$

Two $Q$ network critics are used to prevent value function overestimation. Details on the double $Q$ network, policy updates, and target network mechanisms are in [26] and not discussed here due to space limitations. The SAC algorithm pseudocode for delayed SFDIAs is presented in Table II.

TABLE II PSEUDOCODE OF THE SAC ALGORITHM FOR DELAYED SFDIAS

| |
|---|
| Initialize policy parameters $\phi$, double $Q$-value function parameters $\theta_1, \theta_2$ and the target network parameters $\bar{\theta}_1, \bar{\theta}_2$ with $\theta_1, \theta_2$. |
| Initialize experience replay memory $D$ and BDD thresholds. |
| **while** not converged |
|    **for** each episode **do** |
|       Randomly select a start point in the training dataset and obtain the initial state $s_0$. |
|       **while** not *done* |
|          Select action $a_t$ based on state $s_t$ using the policy. |
|          Input action $a_t$ to environment, acquire *done* signal, reward $r_t$ and next state $s_{t+1}$. |
|          Memorize $(s_t, a_t, r_t, s_{t+1}, done)$ in experience replay buffer $D$. |
|       **end** |
|    **end** |
|    **for** each gradient step **do** |
|       Randomly sample a minibatch of transitions from $D$. |
|       Update the parameters of the $Q$-function, the policy network, and the target network. |
|    **end** |
| **end** |

IV. CASE STUDY

In this paper, we implement the delayed SFDIA scheme using the BEMS framework from [28]. This setup features a grid-forming BESS with a capacity of 3 MW/12 MWh and a 4.5 MW PV farm powering an islanded microgrid. The SoC of the BESS is critical for microgrid operational planning within the BEMS, with the SoC profile over a day depicted in Fig. 2. The BEMS confines the SoC to an operational range of 20% to 90%. When the SoC approaches 20%, the BEMS initiates non-critical load shedding to maintain power only for critical loads. Conversely, exceeding 90% prompts load engagement or PV power curtailment. Insufficient PV power to recharge the BESS below 20% results in shutdown of the BESS and the microgrid.

### A. Simulation Model, Dataset and HIL Testbed

We tested the proposed attack scheme on a centralized distribution model with 5 buses, based on the IEEE 123 bus system, as illustrated in Fig. 1. All the feeder loads of the IEEE-123 bus system in [28] were aggregated and considered as a centralized load connected to Bus 5. Load power consumption data was sourced from actual residential users from Austin, TX. The PV farm and BESS model developed in [29]-[33] were used, with the irradiance profile coming from the actual PV data in North Carolina. The BESS utilizes one RC-branch battery model with parameters summarized in Table III [34], [35].

Table III. Battery model parameters

| Parameters | Values |
|---|---|
| Nominal capacity and power | 12 MWh/6667 Ah, 3 MW |
| Nominal DC voltage and current | 1800V, 1667A |
| DC voltage range | 1607 ~ 2100V |
| DC current range | -1867A ~ 1867 A |
| $R_0$ (per cell) | 1.3 m$\Omega$ |
| $R_1$, $C_1$ (per cell) | 4.2 m$\Omega$, 17111 F |
| Cells (series*parallel) | 492*98 |

To replicate a practical ADN with communication capabilities, we developed a real-time simulation testbed on the OPAL-RT platform. As shown in Fig.1, the distribution network, loads, PV farm, and BESS were modeled on the OPAL-RT platform, while the ADN control center operated on a separate PC. A relay node between the OPAL-RT system and the ADN control center used the Modbus communication protocol to gather measurements and transmits them via TCP/IP. Measurement data was collected every minute for SE, with commands issued every 15 minutes.

We conducted a 20-day simulation, generating measurement noises randomly using normal distributions with zero means and predefined standard deviations: 1% for real-time magnitude, 0.5% for phasor measurements, and 2% for power measurements [11]. The SCADA measurements $z_{SCADA}$ included voltage and current phasor at Bus 2, power injection of all nodes, and power flow of the 4 lines in the distribution network. Additionally, the BESS measurement $z_{BESS}$ included the battery SoC, DC voltage and current, and the modulation index of the BESS inverter. There were collected and sent to the ADN control center along with the network measurements.

### B. Offline training

Assuming the attacker lacks access to the BEMS, we use the first 18 days of historical operation data (1-minute resolution, 25920 data points) for offline training, and the last 2 days for online testing. Both the actor and critic functions in the DRL are modeled using fully connected neural networks. The deep learning framework is implemented in PyTorch, and the algorithm is trained on an NVIDIA RTX 3080 GPU. Table IV summarizes the training hyperparameters.

Table V details the reward parameters. During training, attacks are initiated at random hours, with the BDD trigger threshold set at 99% of the maximum residual error during normal operations. Each episode lasts 10 hours (600 steps) for both unconstrained and constrained SoC error attacks, with $t_d$ set at the 7.7[th] hour for the latter. The target error in the constrained attack is randomly selected from 5% to 30% in 5% intervals. The agent is trained to maximize the SoC error or achieve the target SoC error within the attack duration. If falsified data violates any BDD constraints, a penalty is applied, and the episode ends immediately. SoC operation range constraints are removed during training since offline training uses fixed historical data with actual SoC ranging from 20% to 90%, and the BEMS cannot respond to false SoC data.

Table IV. Hyper parameter for DRL offline training

| Hyper parameter | Values |
|---|---|
| Optimizer | Adam |
| Learning rate | 3e-4 |
| Discount ($\gamma$) | 0.99 |
| Replay buffer size | $10^6$ |
| Number of hidden layers (all networks) | 3 |
| Number of hidden units per layer | 256 |
| Batch size | 256 |
| Target smoothing coefficient ($\tau$) | 0.005 |
| Temperature coefficient ($\alpha$) | 0.5 |

Table V. Reward Parameters

| Parameters | Values | Parameters | Values |
|---|---|---|---|
| $k_{u1}$ | +1 | $k_{t1}$ | 0.2 |
| $k_{u2}$ | 50 | $k_{t2}$ | 2000 |
| $k_{u3}$ | 2500 | $k_{t3}$ | 100 |
| $k_p$ | -500 | | |

In Fig. 5, mean episode reward curves for the unconstrained and constrained SoC error attacks are shown with a 50-epoch sliding window during training. Initially, the average score remains at -500 due to exploration. The BDD constraints are easily violated, resulting in penalties and episode termination. Over training, the reward steadily rises, nearing convergence after 3000 episodes and 9000 episodes in Fig. 5 (a) and (b). The model learns to approach or achieve attack objectives without triggering BDD, driven by the reward function.

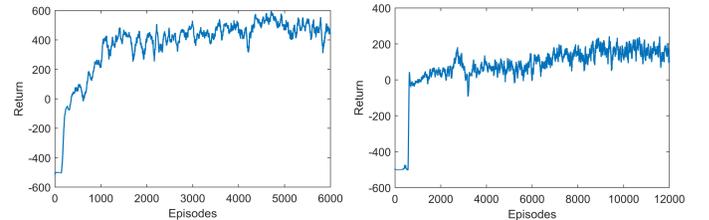

(a) unconstrained SoC error attack    (b) constrained SoC error attack
**Fig. 5.** Training process of the DRL agent.

### C. Offline test

Using historical data from the last two days, we conducted an offline test employing the agent within the BEMS framework [28]. Two attack scenarios were established: one concluding at 1 am with an expected SoC of approximately 40%, and another ending at 8 am with an anticipated SoC near 20%. The attacks aimed to introduce SoC estimation errors, potentially leading to the shutdown of the BESS-supplied system around midnight or between 8 am and 12 pm, with the latter being a critical power-supply period. Table VI

Table VI. Offline attack test results of unconstrained and constrained SoC error attack

| | No attacks | Unconstrained | | | | | | Constrained | | | | | | | | | |
|---|---|---|---|---|---|---|---|---|---|---|---|---|---|---|---|---|---|
| Cases | 0 | 1 | 2 | 3 | 4 | 5 | 6 | 7 | 8 | 9 | 10 | 11 | 12 | 13 | 14 | 15 | 16 |
| $t_s$ (hour) | - | 2nd | 8th | 16th | 2nd | 8th | 16th | 2nd | 8th | 16th | 8th | 8th | 2nd | 8th | 16th | 8th | 8th |
| $t_d$ (hour) | - | - | - | - | - | - | - | 21st (21 pm, 1st day) | | | | | 25th (1 am, 2nd day) | | | | |
| $t_e$ (hour) | - | 25th (1 am, 2nd day) | | | 32nd (8 am, 2nd day) | | | 25th (1 am, 2nd day) | | | | | 32nd (8 am, 2nd day) | | | | |
| $\Delta\Phi^*$(%) | - | - | - | - | - | - | - | 20 | | | 10 | 30 | 20 | | | 10 | 30 |
| $\Delta\Phi_{t_d}$(%) | - | - | - | - | - | - | - | 18.3 | 16.5 | 4.6 | 13.1 | 16.6 | 19.9 | 19.6 | 8.4 | 10.9 | 19.7 |
| $\Delta\Phi_{t_e}$(%) | - | 12.8 | 11.8 | 6.1 | 18.1 | 17.1 | 11.1 | 20.4 | 19 | 7.2 | 11.0 | 18.7 | 20.4 | 20.4 | 16.5 | 10.4 | 27.4 |
| $\bar{r}_{med}$ | 0.062 | 0.073 | 0.073 | 0.081 | 0.074 | 0.073 | 0.073 | 0.075 | 0.078 | 0.078 | 0.074 | 0.077 | 0.075 | 0.078 | 0.080 | 0.074 | 0.080 |

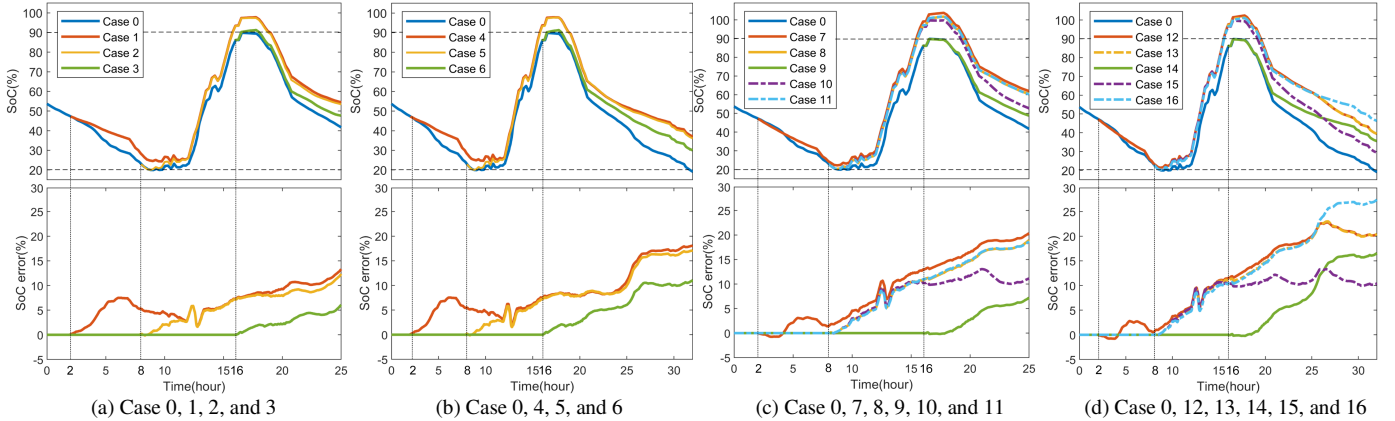

(a) Case 0, 1, 2, and 3    (b) Case 0, 4, 5, and 6    (c) Case 0, 7, 8, 9, 10, and 11    (d) Case 0, 12, 13, 14, 15, and 16

**Fig. 6.** SoC and SoC error profiles of offline test results with different start, end times, or targeted SoC errors: unconstrained – (a), (b); constrained – (c), (d).

summarizes 16 offline attack test cases with varying start times, durations, and targeted SoC errors under constrained attack mode. Figure 6 depicts the SoC and SoC error profiles, while Fig. 7 shows the residual distribution. Notably, Case 4 had the highest SoC error at 18.1% among constrained attacks. Cases 12, 13, and 15 reached and maintained the targeted error within a ±1% SoC tolerance at specified times in unconstrained cases. Observations from the offline tests include:

- *Attack durations:* Longer attack durations or earlier attack start times usually result in larger SoC errors in unconstrained attacks, and higher chances of reaching and maintaining the targeted SoC error at the designated time $t_d$ and the end time $t_e$ for constrained attacks.

- *SoC error injection:* SoC error generally increases over time but exhibits fluctuations due to varying system statuses and BDD limits. When the actual SoC near certain plateaus, the SoC error or its injection rate tends to decrease, particularly when the actual SoC approaches the 20% plateaus over four hours. Attacks initiated at 2 am exhibit similar final SoC errors to those started at 8 am, as the accumulated SoC error before 8 am drops near zero during the 20% plateaus. Additionally, attacks concluding at the 32nd hour (8 am of the second day) show little variation in SoC error when the actual SoC is approaching 20%.

It is also observed that some false SoC values exceed 90%. This occurs because the SoC range constraint is removed for more efficient offline training. Moreover, the SoC error injection patterns under different attack scenarios are quite similar for the same system operation points. Part of the SoC error curves overlap or display similar variation trend across various attacks.

- *Targeted SoC errors*: For constrained attacks, if the targeted SoC error can be reached and maintained at designated times correlates with the target value and attack duration. The goal can only be reached when the targeted SoC error is achievable within the specified attack duration, as shown in cases 10, 12, 13, and 15. Smaller targets are reached sooner and then fluctuate around the target value (cases 10, 15). If the target is not feasible within $t_d$, it may still be achieved at $t_e$ due to the dual bonus mechanism (cases 7 and 8). The final SoC error will try to get close to target if the target is too large, comparing to the specified attack duration (cases 9, 11, 14, and 16).

- *Residual comparison*: The residuals for all attacked cases are below the threshold and mostly fall within the same range as non-attacked cases. However, the median residual is higher in the attacked cases due to the injected attack vector. Longer attack durations generally result in smaller median residuals for injecting the same SoC error. For unconstrained attacks, case 3 has the highest residual median due to the shortest attack duration. For constrained attacks, both shorter attack durations and larger targeted SoC errors can increase the residual median.

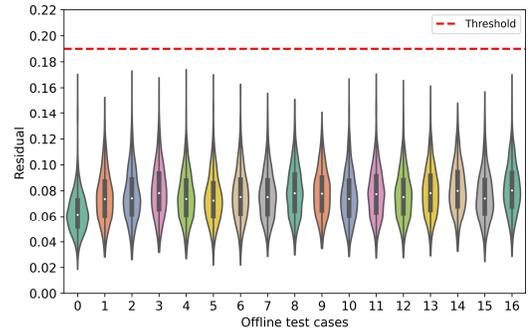

**Fig. 7.** Residual distribution of unconstrained and constrained SoC error attack offline test results.

Overall, it can be observed that the proposed method leverages measurement redundancy and BDD threshold margins to inject SoC errors. The residuals in the attack cases remain within the threshold range but exhibit higher median values because only the battery voltage and current are modified. If a sufficient number of measurements can be altered, the residuals could remain consistent. The proposed methodology is adaptable to such scenarios by incorporating these additional measurements into the action space.

Table VII. Online test results of unconstrained and constrained SoC error attack

| | No attacks | Unconstrained | | | | | | Constrained | | | | | | | | |
|---|---|---|---|---|---|---|---|---|---|---|---|---|---|---|---|---|
| Cases | 0 | 1 | 2 | 3 | 4 | 5 | 6 | 7 | 8 | 9 | 10 | 11 | 12 | 13 | 14 | 15 | 16 |
| $t_s$ (hour) | - | $2^{nd}$ | $8^{th}$ | $16^{th}$ | $2^{nd}$ | $8^{th}$ | $16^{th}$ | $2^{nd}$ | $8^{th}$ | $16^{th}$ | $8^{th}$ | $8^{th}$ | $2^{nd}$ | $8^{th}$ | $16^{th}$ | $8^{th}$ | $8^{th}$ |
| $t_d$ (hour) | - | - | - | - | - | - | - | $21^{st}$ (21 pm, $1^{st}$ day) | | | | | $25^{th}$ (1 am, $2^{nd}$ day) | | | | |
| $t_e$ (hour) | - | $25^{th}$ (1 am, $2^{nd}$ day) | | | $32^{nd}$ (8 am, $2^{nd}$ day) | | | $25^{th}$ (1 am, $2^{nd}$ day) | | | | | $32^{nd}$ (8 am, $2^{nd}$ day) | | | | |
| $\Delta\Phi^*$(%) | - | - | - | - | - | - | - | 20 | | 10 | 30 | | 20 | | | 10 | 30 |
| $\Delta\Phi_{t_d}$(%) | - | - | - | - | - | - | - | 19.9 | 15 | 6.1 | 11 | 15.7 | 19.8 | 19.8 | 11.8 | 10.6 | 20.3 |
| $\Delta\Phi_{t_e}$(%) | - | 14.7 | 13.4 | 8.1 | 17.9 | 15.5 | 9.2 | 20.4 | 19.1 | 11.9 | 10.4 | 19.4 | 21 | 21.1 | 17.9 | 11.6 | 26.2 |
| $\bar{r}_{med}$ | 0.062 | 0.074 | 0.074 | 0.081 | 0.074 | 0.072 | 0.076 | 0.075 | 0.076 | 0.081 | 0.075 | 0.078 | 0.072 | 0.075 | 0.079 | 0.074 | 0.079 |

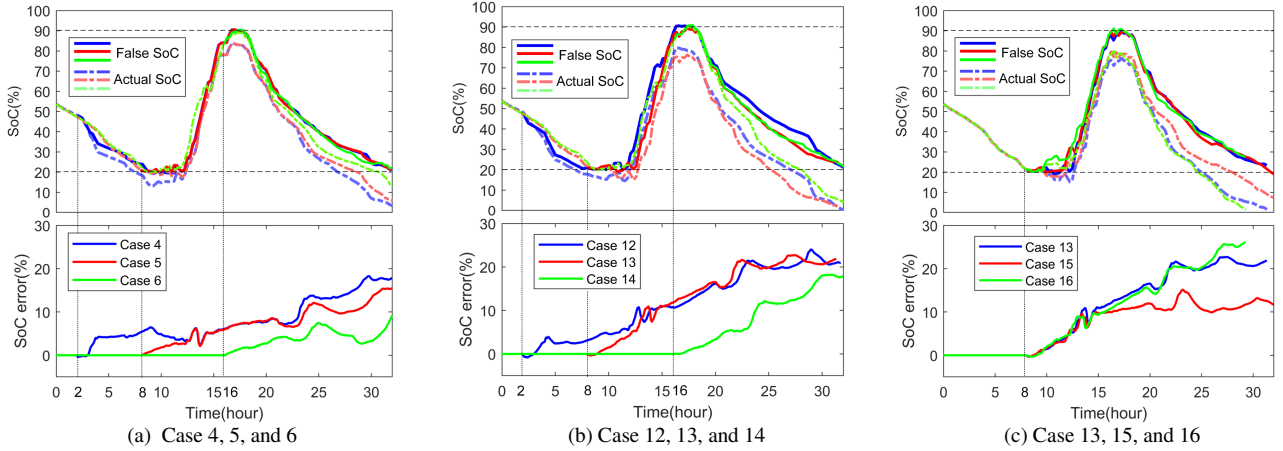

(a) Case 4, 5, and 6          (b) Case 12, 13, and 14          (c) Case 13, 15, and 16

**Fig. 8.** SoC and SoC error profiles of online test results with different start, end times, or targeted SoC errors: unconstrained – (a); constrained – (b), (c).

*D. Online Test*

After the offline training and testing, the agent is applied to online testing using the same load and PV profile from the offline test. For the online test, the agent is implemented for real-time attack, transmitting the false battery voltage, current and SoC to the ADN control center. The attack is done to mislead the BEMS into making inappropriate energy decisions. The simulation results of all online test cases are summarized in Table VII.

By comparing Table VI and VII, it is evident that the online attack result is highly similar to those of the offline attacks. In the unconstrained attack scenario, case 4 shows the highest SoC error due to the longest attack duration. For constrained attacks, a longer attack duration increases the likelihood of achieving the targeted SoC error at two specified times. Figure 8 illustrates part of the false and actual SoC profiles, as well as the SoC error profiles of the online test cases. The primary distinction between offline and online attacks lies in how the actual SoC is affected. During offline attacks, the actual SoC remains within the range of 20% to 90% and is not influenced by the false SoC, as shown in Fig. 6. In contrast, during online attacks, the false SoC is treated as the actual SoC by the BEMS and is regulated within the desired range. As shown in Fig. 8, all false SoC is within the range of 20% to 90% but with slightly different shapes, even though the bounded-range BDD for SoC is disabled during offline training. Consequently, the actual SoC is impacted by the false SoC or the injected SoC error.

For unconstrained attacks, all cases introduced SoC errors at the desired end times (1 am and 8 am). Case 1 injected the largest SoC error at 1 am, causing the energy deficiency for the system's operation throughout the night. If the attack ends at this time, the BEMS will detect the actual SoC value. This could shut down the system when the actual SoC is below 20% and there is no PV to charge BESS or provide power for loads during the night. Case 4 injected the largest SoC error of 17.9% when the actual SoC is only 3.2% at 8 am. Since 8 am to 12 am is designated as a critical power supply period in the BEMS, nearly all PV power is used to supply loads while the BESS primarily provides voltage and frequency support, keeping the BESS SoC around 20% during this period. If the attack ends at 8 am, the system will shut down due to the loss of voltage and frequency support, disrupting the power supply during critical periods.

Similarly, for constrained attacks, the same system shutdown will happen due to the injected SoC estimation error. Comparing to the unconstrained attack, the injected SoC error in constrained attacks could be higher and more deliberate, resulting in longer periods of system shutdown. It is also observed that the actual SoC drops to around 0 before the attack end time in cases 12, 13 and 16.

Figure 9 presents the residual distribution for all online teste cases. All residuals remain below the BDD threshold. Compared to the non-attack case, the median residuals in the attacked cases are higher, with cases 3, 9, 11, 14 and 16 showing the highest median residuals. This is because all cases aim to reach a large SoC error target within a limited attack duration.

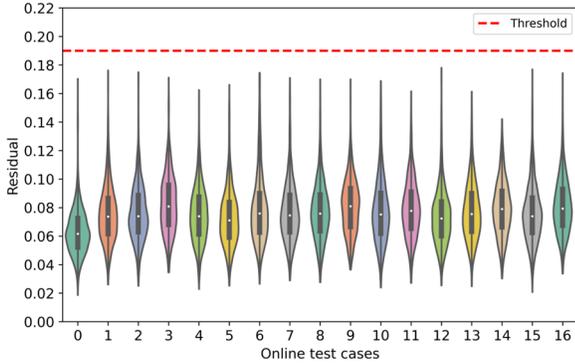

**Fig. 9.** Residual distribution of unconstrained and constrained SoC error attack online test results.

## V. Conclusion

To address the substantial computational demands of nonlinear delayed SFDIAs while ensuring real-time online deployment, we proposed a DRL-based delayed SFDIA algorithm specifically designed to disrupt BESS operations during targeted time periods. This innovative algorithm degrades the SoC estimation of BESS over time by strategically altering battery voltage and current. Our method exploits measurement redundancy and BDD threshold margins to effectively inject potential SoC errors.

The RL agent, through interaction with the ADN environment incorporating three distinct BDD algorithms, learns to generate a sequence of attack vectors for delayed SFDIA attacks. These vectors are capable of evading BDD detection, successfully introducing the desired SoC error by the end of the attack period. We introduced two distinct attack modes: unconstrained and constrained SoC error attacks. The constrained mode allows for precise control of the injected SoC error, maintaining it within a targeted range, while the unconstrained mode aims to generate the largest possible errors.

Our proposed attack methodology and strategies were rigorously tested using a HIL platform. The results demonstrated the effectiveness of our approach in injecting the desired SoC error without triggering any BDD mechanisms. This confirms the potential of our DRL-based delayed SFDIA algorithm to meet real-time deployment requirements while effectively compromising BESS operations. The success of this method underscores the need for enhanced BDD mechanisms to counteract sophisticated SFDIA threats and safeguard BESS integrity.